\documentclass{esub2acm}
 \usepackage{amssymb}

 \newcommand{\be}{\begin{equation}}
 \newcommand{\ee}{\end{equation}}
 \newcommand{\ea}{\end{array}}

\begin{document}

 \title{Algorithm xxx: Modified Bessel functions of imaginary order and
 positive argument}
 \author{Amparo Gil \footnote{Present address:  Departamento de Matem\'aticas, 
 Estad\'{\i}stica y Computaci\'on. 
 U. de Cantabria, 39005-Santander, Spain}\\
         Departamento de Matem\'aticas, U. Aut\'onoma de Madrid, 
         28049-Madrid, Spain\\
         \\
     Javier Segura\\
         Departamento de Matem\'aticas, Estad\'{\i}stica y 
 Computaci\'on. 
 U. de Cantabria, 39005-Santander, Spain\\   
     \and
     Nico M. Temme\\
     CWI, P.O. Box 94079, 1090 GB Amsterdam, The Netherlands}

 \begin{abstract}
 Fortran 77 programs for the computation of modified Bessel functions of purely
imaginary order are presented. The codes compute 
the functions $K_{ia}(x)$, $L_{ia}(x)$ and their derivatives for real $a$ 
and positive $x$;
these functions are independent solutions of the differential equation 
 $x^2 w'' +x w' +(a^2 -x^2)w=0$. 
 The code also computes exponentially scaled functions.
 The range of computation is $(x,a)\in (0,1500]\times
[-1500,1500]$ when scaled functions are considered and it is larger
than $(0,500]\times [-400,400]$ for standard IEEE double precision arithmetic. 
The relative accuracy is better than $10^{-13}$ in the range 
$(0,200]\times [-200,200]$ and close to $10^{-12}$ in $(0,1500]\times [-1500,1500]$.
 \end{abstract}
 % A category with only the three required fields
 %A category including the fourth, optional field
 %\category{H.32.m} {NonInformation Systems} {Miscellaneous}
 %[SGML Database Systems]
 \begin{bottomstuff}

 % You might add a line or two here if a version of your article appeared
 % previously, or if the work was supported by an organization or conducted
 % under a grant.
 %\begin{authinfo}
 %\name{A. Gil} 
 %\address{P.O. Box 1221, Dublin, OH 43017-6221}
 %\affiliation{Kassel University, Mathematics Department} 
 %use if there are several authors with different affiliations
 %\biography{} optional; not generally used.
 %\end{authinfo}
 % Here's a neat thing:  all you have to put in your document is the command
 %"\permission" and LaTeX automatically enters the complete text of the
 %"Permission to copy..." boilerplate
 %\permission
 \end{bottomstuff}
 \markboth{Amparo Gil, Javier Segura and Nico M. Temme}
      {Modified Bessel functions of imaginary orders}
 \maketitle
 \section{Introduction}

 These routines compute the functions $K_{ia}(x)$ and $L_{ia}(x)$,
which constitute a numerically satisfactory pair of independent
solutions of the modified Bessel equation of imaginary order:

\begin{equation}
x^2 w'' +x w' +(a^2 -x^2)w=0.
\label{ODE}
\end{equation}

The routine DKIA computes $K_{ia}(x)$ and its derivative. DLIA 
computes $L_{ia}(x)$ and its derivative. Both routines require
that a method for computing Airy functions of real variable is
available. The routines use Algorithm 819 \cite{Gil3}.

The algorithm combines different methods of evaluation in
different regions. Comparison between the different methods 
(see accompanying paper \cite{Gil0}), together with the use
of the Wronskian relation, has been used to determine the
accuracy of the algorithm (see Section \ref{sec2}).

The algorithm computes either the
functions $K_{ia}(x)$, $L_{ia}(x)$ and their derivatives or scaled functions 
which can be
used in larger regions than the unscaled functions. The scaled functions are
defined in the accompanying paper \cite{Gil0} (Section 3, Eqs. (43) and (44)) and in the
comment lines of the algorithms.
The relative accuracy of the algorithms is (except close to the zeros of the functions)
better than $10^{-13}$ in the $(x,a)$-region $D_1\equiv(0,200] \times [-200,200]$, better than 
$5\times 10^{-13}$ in $D_2 \equiv (0,500] \times [-500,500]$ and close to $10^{-12}$ in 
 $D_3 \equiv (0,1500] \times [-1500,1500]$.
Both scaled
and unscaled functions can be computed in $D_1$ but only scaled functions
can be evaluated in the whole domains $D_2$ and $D_3$ without causing
overflows/underflows for typical IEEE machines. 
See the accompanying paper \cite{Gil0}, Section 3.

\section{Regions of application of the methods and accuracy.}
\label{sec2}

 Because the scaled and unscaled functions are even functions of $a$, we can restrict the
study to $a \ge 0$.

 In order to determine the region of applicability of each method, we have compared all
 the methods available with the non-oscillating integral representations in the 
accompanying paper \cite{Gil0}, Section 2.4,
 which are expected to be valid in all the $(x,a)$ plane except close to
 $a=x$. This is so because, as explained in \cite{Gil1}, the integration paths become 
 increasingly
 non-smooth as one approaches the transition line $a=x$ (in this region, the uniform asymptotic
 expansions in the accompanying paper, Section 2.3, 
are the best option). The remaining approaches have
 little or no overlap in their regions of validity, except for the continued fraction
 method for the computation of $K_{ia}(x)$ and $K_{ia}' (x)$ and the asymptotic expansions
 for large $x$; in fact, the validity region of the CF-method completely covers that 
 of the asymptotic expansion; 
therefore, the CF method is preferred over the asymptotic expansion
for large $x$ for the computation of $K_{ia}(x)$ and its derivative.

 As we will next describe, for not too large values of $a$ and $x$ there are 
at least two alternative methods of computation in any region of the $(x,a)$ plane. 
This is important in order to determine the accuracy reachable by the algorithm. A second
validation of the methods is provided by the Wronskian relation 
$$
K_{ia}(x)L_{ia}'(x)-K_{ia}'(x)L_{ia}(x) =1/x,
$$
\noindent which is also satisfied by the scaled functions $\widetilde{K_{ia}}(x)$, 
$\widetilde{K_{ia}'}(x)$, $\widetilde{L_{ia}}(x)$, 
$\widetilde{L_{ia}'}(x)$.

Of course, when two different approaches are available within a given accuracy in a same
region, we choose the fastest method of computation. By order of speed, from fastest to
slowest we can list the different methods of computation described in \cite{Gil0} as follows:

\begin{enumerate}
\item{Series expansions and continued fraction method.}
\item{Asymptotic expansions.}
\item{Integral representations.}
\end{enumerate}

Figures 1,2 show the regions in the plane $(x,a)$ where 
the different methods are used 
in the routines DKIA and DLIA, respectively. In the first
comment lines of the codes, the explicit equations for the
curves separating the different regions are given.

\vspace*{0.1cm}
 \centerline{\protect\hbox{\psfig{file=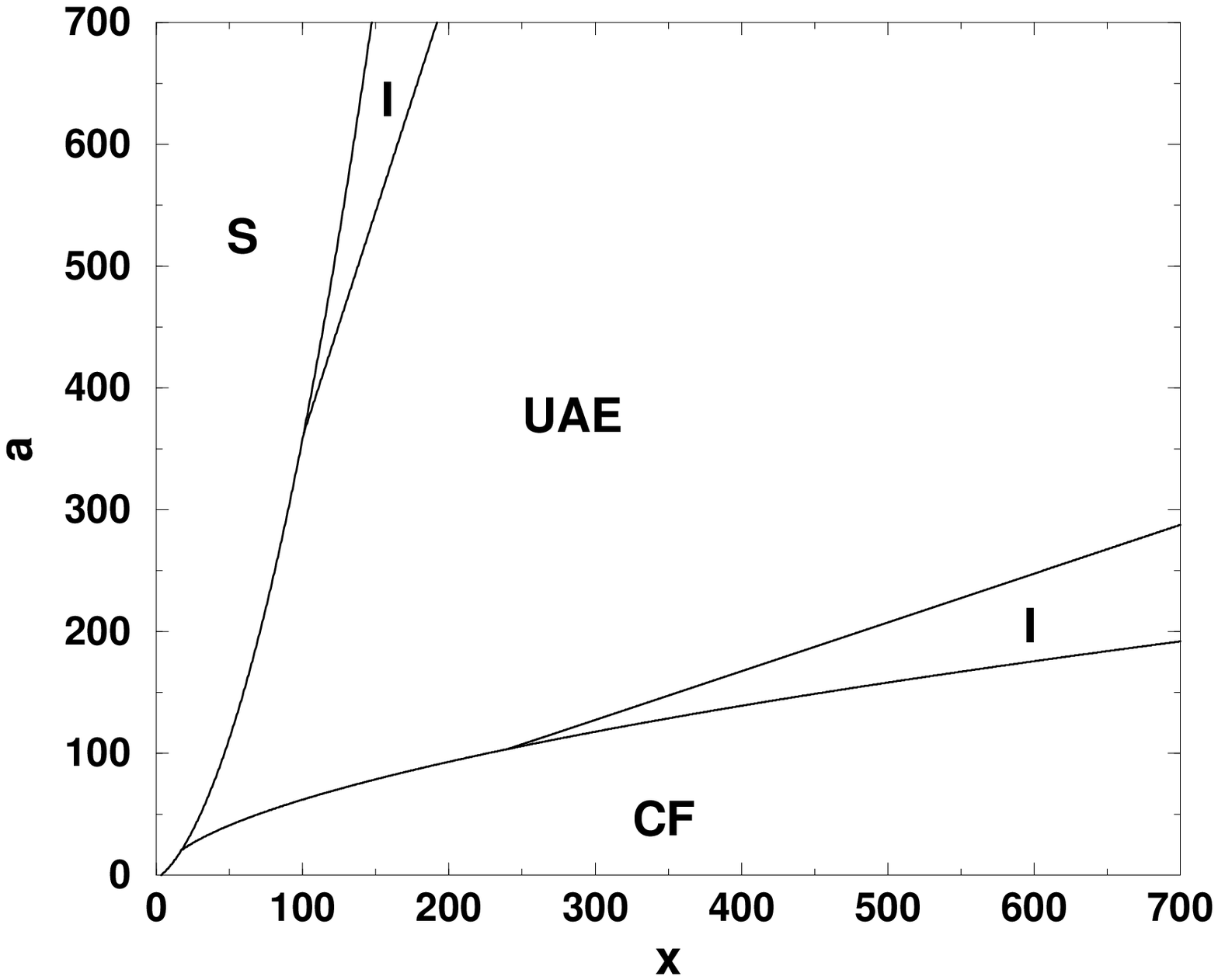,width=6cm}}}
 \vspace*{0.1cm}
 {\bf Figure 1.} {\footnotesize Regions of applicability of each method for the
$K_{ia}(x)$, $K'_{ia}(x)$ functions.  {\bf UAE}: uniform Airy-type asymptotic expansion;
{\bf I}: integral representations; {\bf S}: power series; {\bf CF}: for continued fraction and 
{\bf AE}: asymptotic expansions.}

\vspace*{0.1cm}
\centerline{\protect\hbox{\psfig{file=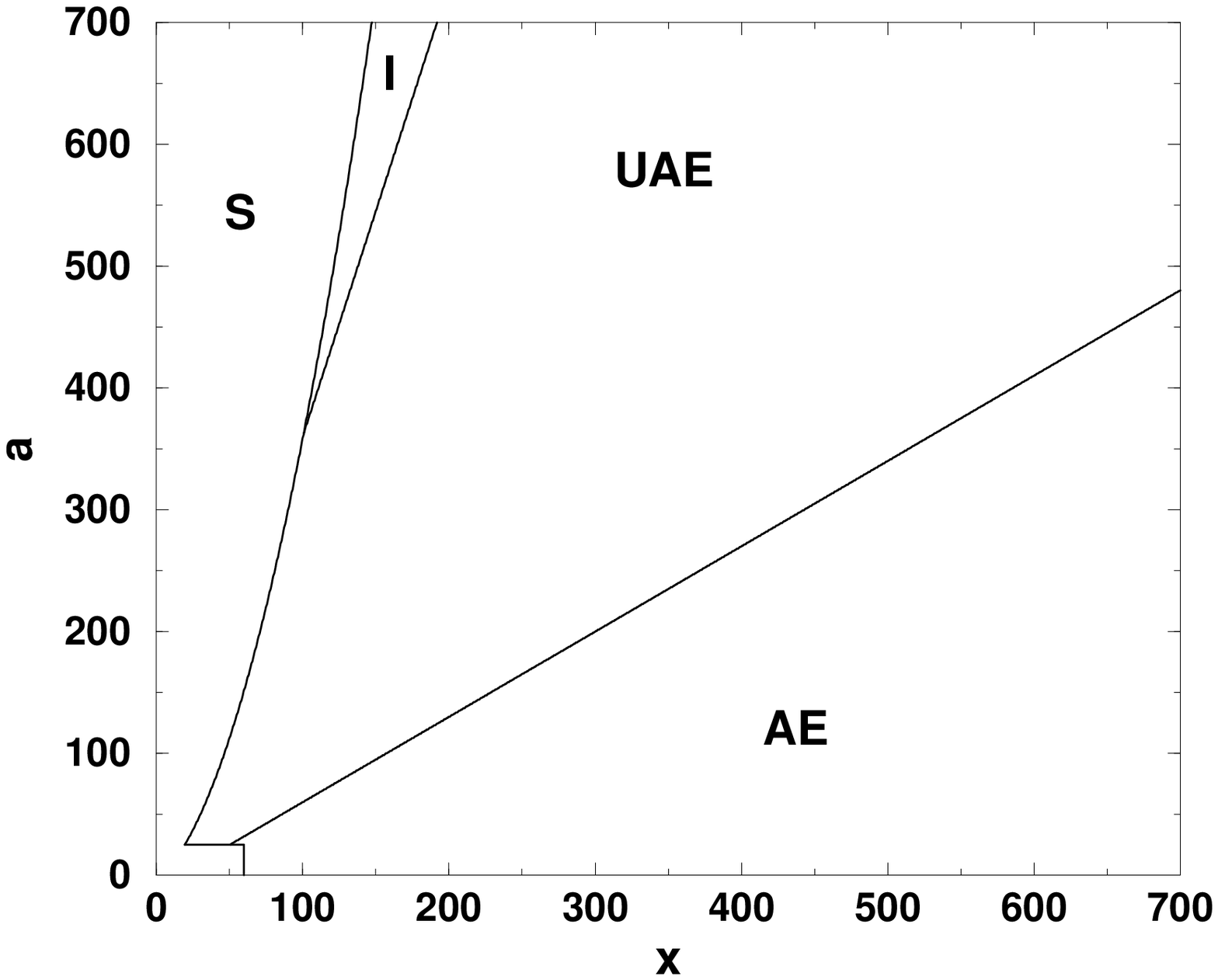,width=6cm}}}

{\bf Figure 2.} {\footnotesize Regions of applicability of each method for the
$L_{ia}(x)$, $L'_{ia}(x)$ functions. {\bf UAE}: uniform Airy-type asymptotic expansion;
{\bf I}: integral representations; {\bf S}: power series; {\bf CF}: for continued fraction and 
{\bf AE}: asymptotic expansions.}

\vspace*{0.1cm}
Both the direct comparison with integral representations as well as the Wronskian check
show that the unscaled and scaled functions can be computed with an accuracy better
than $10^{-13}$ in the region $(x,a)\in (0,200]\times [0,200]$
 without using integral representations. This accuracy was found to be in a good compromise
with efficiency. We expect that function values outside this
range will not be needed very often.
As the range considered is increased, the accuracy decreases
 mildly, being better than $5\,\times 10^{-13}$ in $(0,500]
\times [0,500]$ and
close to $10^{-12}$ in the full range of
computation $(0,1500]\times [0,1500]$. Of course, near the zeros of the functions (there are infinitely
many of them in the oscillatory region $x<a$) relative accuracy loses meaning and
only absolute accuracy makes sense.

\section{Timing}

Figure 3 shows the CPU time spent by the code in the regions
$(x,a)\in (0,L]\times [0,L]$; these CPU times refer to a Pentium-II
350-Mhz processor running under Debian-Linux; the compiler used was
the GNU Fortran compiler g77.

 \vspace*{0.1cm}
 \centerline{\protect\hbox{\psfig{file=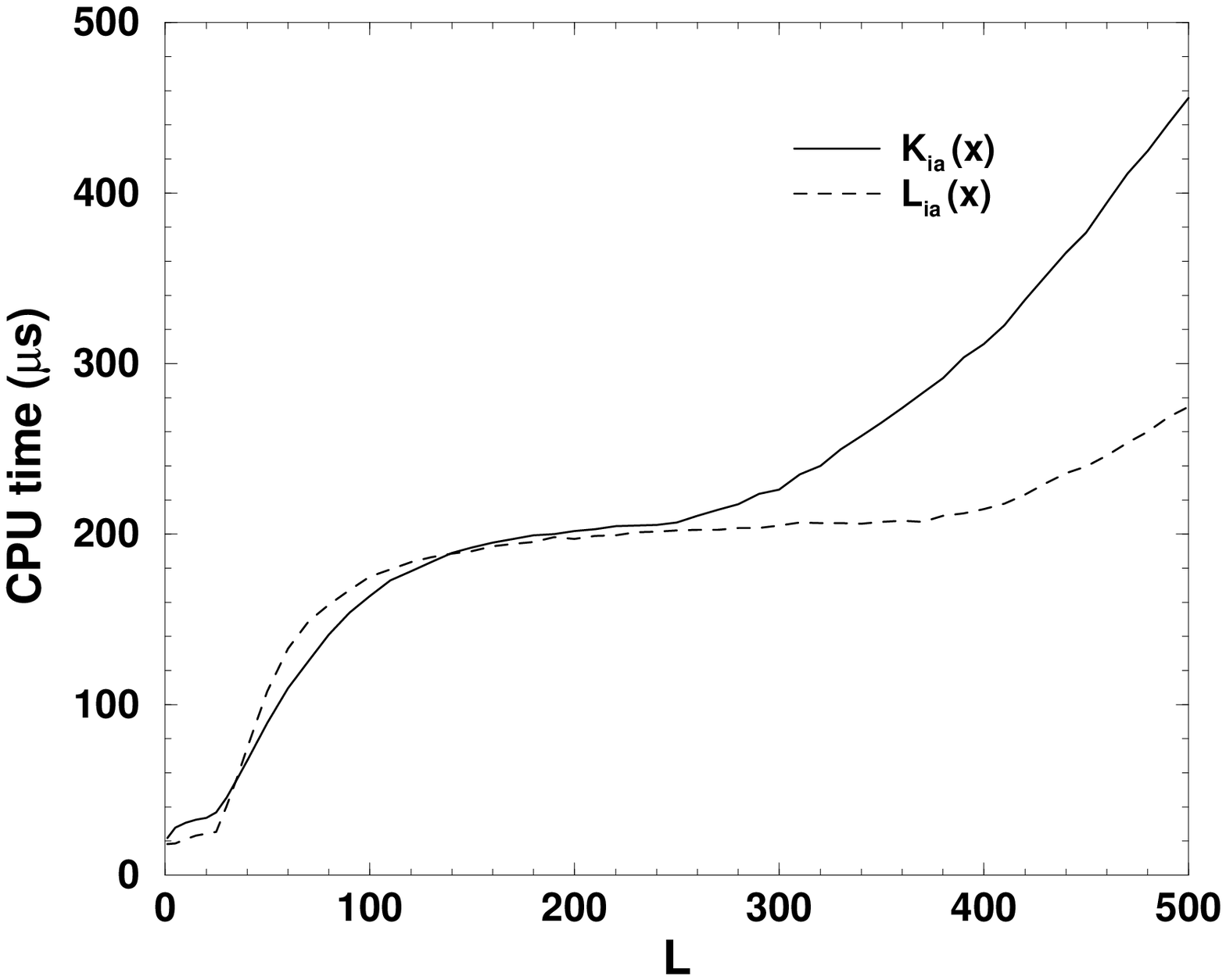,width=6cm}}}
 \vspace*{0.1cm}

{\bf Figure 3.} {\footnotesize CPU times in $(0,L]\times [0,L]$ as
a function of $L$.}
\vspace*{0.1cm}

It is apparent from the figure 
that different methods
with different speeds appear as $L$ grows. For small $L$, only 
series (and the CF method for $K_{ia} (x)$) are needed, which are the 
fastest methods. As soon as asymptotic expansions come into play 
the CPU times tend to increase more rapidly. For $L$ larger than $200$
the effect of the use of quadratures becomes prominent.

Tables 1 and 2 show the average time spent by each method of computation 
in microseconds,
both for DKIA and DLIA routines. Also, the percentage of use of each method
is shown. These data are shown for two different regions of computation:
the ``$10^{-13}$ accuracy region'' ($(0,200]\times [0,200]$) (Table 1) 
and the full
range permitted by the routines ($(0,1500]\times [0,1500]$) (Table 2).

$$
\begin{array}{c|ccccc}
 & \mbox{S} & \mbox{CF} & \mbox{AE} & \mbox{UAE} & \mbox{I} \\ 
\hline
\% \mbox{ of use}        &      &     &  &       &   \\        
  \mbox{DKIA}  & 22   & 29  &  & 49  & 0 \\
   \mbox{DLIA}  & 25  &   & 28.4  & 46.6  & 0 \\
\hline
\mbox{Average time } (\mu s)   &     &      &  &      &  \\    
    \mbox{DKIA}    & 51  & 55   &  & 370  & 0 \\
     \mbox{DLIA}   & 44  &   & 91  & 352  & 0 \\
\end{array}
$$
{\bf Table 1.} {\footnotesize Percentage of use and average CPU times 
of each method
in the region $(0,200]\times [0,200]$.}

\vspace{0.1cm}
$$
\begin{array}{c|ccccc}
 & \mbox{S} & \mbox{CF} & \mbox{AE} & \mbox{UAE} & \mbox{I} \\ 
\hline
\% \mbox{ of use}        &      &     &  &       &   \\        
  \mbox{DKIA}  & 9.7   & 12.6  &  & 65.6  & 12.1 \\
   \mbox{DLIA}  & 9.7  &   & 34.3  & 51.7  & 4.3 \\
\hline
\mbox{Average time } (\mu s)   &     &      &  &      &  \\    
    \mbox{DKIA}    & 79  & 91   &  & 178  & 12130 \\
     \mbox{DLIA}   & 61  &   & 284  & 279  & 19475 \\
\end{array}
$$
{\bf Table 2.} {\footnotesize Percentage of use and average CPU times 
of each method
in the region $(0,1500]\times [0,1500]$.}

\vspace{0.1cm}
The algorithm has been tested in several computers and operating
systems (Pentium II PC under Debian Linux, Pentium IV laptop under
Windows XP, Pentium IV PC under Windows XP and Red Hat Linux and 
Solaris 8 workstation) and several compilers (g77 and f77 for 
Linux and Windows, Digital Fortan and Lahey Fortran for Windows 
and the Sun Fortran 95).

\section{Comparison with existing software}

In spite of the many applications of the modified Bessel functions
of imaginary order, the only previously published algorithm,
 as far as the authors know, is the code by 
 Thompson and Barnett \cite{Tho} (TB algorithm from now on). 
 This code is intended for the computation
 of Coulomb wave functions of complex order and arguments, which have 
 as a subset modified Bessel functions of complex orders. 
 It is not surprising that, as we later discuss, the TB algorithm has limitations
 when computing the $K_{ia}(x)$ and $L_{ia}(x)$ functions:
 this is a more general algorithm than ours and it was specially designed
for real parameters, although it can be applied for complex parameters 
too. Apart from this code, there are no available algorithms in the public domain, 
 as can be judged by browsing at GAMS 
(Guide to Available Mathematical Software: 
 http://gams.nist.org), which are able to compute 
$K_{ia}(x)$, $L_{ia}(x)$ and their derivatives.

 As Figure 4 shows, the TB algorithm (program COULCC available from
 CPC library) is inaccurate for
 imaginary orders and suffers from overflow and convergence problems.
 We have tested the program COULCC \cite{Tho} against our algorithms.
 Figure 4 shows the comparison. At the lighter points the flag for detecting
 errors in COULCC was different from zero, showing that 
 the algorithm 
 failed to converge or suffered overflows. The rest of points show 
 discrepancies with our code, which persist for $10^{-8}$ accuracy
 and lower. Of course, at the lighter points all the accuracy is lost and
 the problems persist for larger values of $x$ and $a$.

\vspace*{0.5cm}
\begin{minipage}{6cm}
\centerline{\protect\hbox{\psfig{file=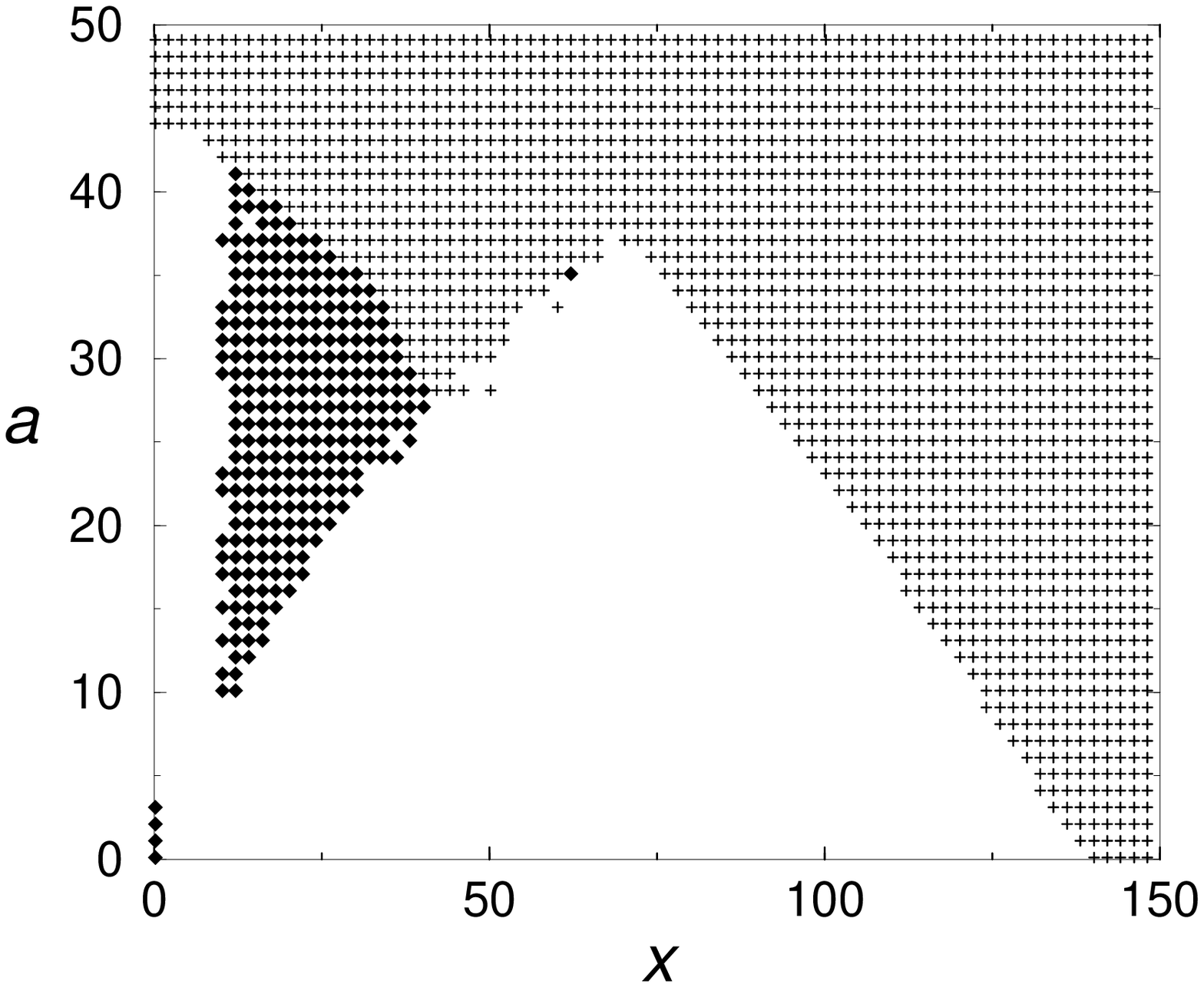,width=5.5cm}}}
\end{minipage}
\hfill
\begin{minipage}{6cm}
\centerline{\protect\hbox{\psfig{file=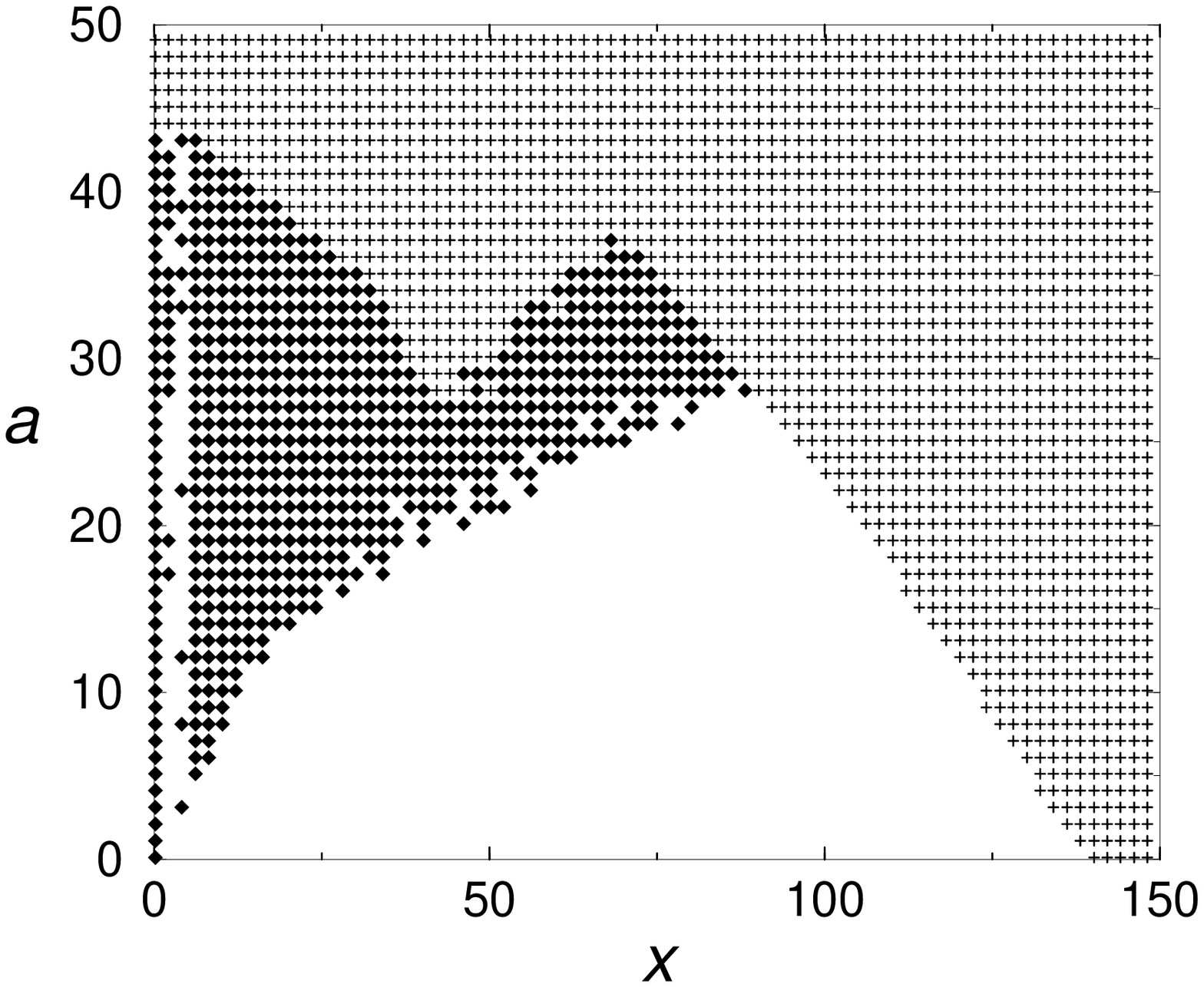,width=5.5cm}}}
\end{minipage}
 {\bf Figure 4.} {\footnotesize Regions where the TB algorithm 
fails due to poor convergence or overflows (light shaded points)
and points where accuracy is lost (dark shaded points). The
discrepancy is greater than $10^{-8}$ in the left figure and
 greater than $10^{-13}$ in the right figure.}

\vspace*{0.1cm}
Therefore, our algorithm seems to be the first accurate and efficient code
which is capable of computing modified Bessel functions of imaginary order
with an accuracy close to full double precision in a wide domain.

 %ACKNOWLEDGEMENTS are optional
 \begin{acks}
 A. Gil acknowledges financial support from   Ministerio de 
 Ciencia y Tecnolog\'{\i}a (Programa Ram\'on y Cajal). 
 A. Gil and J. Segura acknowledge CWI 
 Amsterdam for the hospitality and financial support.  
 \end{acks}
 %APPENDICES are optional
 %\appendix
 %Appendix A

 %\bibliographystyle{esub2acm}
 %\bibliography{sampdoc2e}
 
 % Note! You'll need to *import* the .bbl file INTO your master file
 % prior to submitting it to ACM.
 %
 %
 \end{document}